\documentclass[preprint2]{aastex}

\usepackage{epsf}
\usepackage{graphics}
\usepackage{color}

\newcommand{\be}{\begin{equation}}
\newcommand{\ee}{\end{equation}}

\newcommand{\ax}{$\alpha_{\rm X}$}

\newcommand{\aox}{$\alpha_{\rm ox}$}

\newcommand{\rb}[1]{\raisebox{1.5ex}[-1.5ex]{#1}}

\newcommand{\plm}{$\pm$}
\newcommand{\nh}{$N_{\rm H}$}
\newcommand{\swift}{{\it Swift}}
\newcommand{\chandra}{{\it Chandra}}
\newcommand{\wpvs}{{WPVS~007}}

\shorttitle{Swift observation of WPVS 007}
\shortauthors{Grupe et al.}

\begin{document}


\def\etal{{\it et\thinspace al.}\ }
\def\alp{{$\alpha$}\ }
\def\al2{{$\alpha^2$}\ }

%
%
%


\title{An Update on the X-ray transient Narrow-Line Seyfert 1 galaxy 
WPVS 007:  \swift\ observations of UV variability and persistence of X-ray
faintness
}


\author{Dirk Grupe\altaffilmark{1}
\email{grupe@astro.psu.edu},
Patricia Schady\altaffilmark{1,2},
Karen M. Leighly\altaffilmark{3},
Stefanie Komossa\altaffilmark{4},
Paul T. O'Brien\altaffilmark{5},
John A. Nousek\altaffilmark{1}
}

\altaffiltext{1}{Department of Astronomy and Astrophysics, Pennsylvania State
University, 525 Davey Lab, University Park, PA 16802} 

\altaffiltext{2}{Mullard Space Science Laboratory, Holmbury St. Mary, Dorking,
Surrey RH5 6NT, U.K.; 
email: ps@mssl.ucl.ac.uk}

\altaffiltext{3}{Homer L. Dodge
Department of Physics and Astronomy, University of Oklahoma, 
440 West Brooks Street, Norman, OK 73019; email: leighly@nhn.ou.edu}

\altaffiltext{4}{Max-Planck-Institut f\"ur extraterrestrische Physik, Giessenbachstr., D-85748 Garching,
Germany; email: skomossa@mpe.mpg.de}

\altaffiltext{5}{Department of Physics \& Astronomy, University of Leicester,
Leicester, LE1 7R, UK, email: pto@star.le.ac.uk}




\begin{abstract}
We report on the detection of UV variability and the persistence of X-ray
faintness of the X-ray transient
Narrow-Line Seyfert 1 galaxy WPVS 007 based on the first year
of monitoring this AGN with \swift\ between 2005 October and 2007 January.
WPVS 007 has been an unusual source. While being
X-ray bright during the ROSAT All-Sky Survey it has been extremely faint in all
following X-ray observations.
\swift\ also finds this NLS1 to be X-ray faint and not detected in the
\swift~X-Ray Telescope at an 3$\sigma$ upper limit of $2.6\times 10^{-17}$ W
m$^{-2}$  in the 0.3-10.0 keV band
and confirms that the AGN is still in a low state.
During the 2006 July and December observations with \swift's UV-Optical Telescope
(UVOT)
the AGN became fainter by about 0.2 mag in the UV filters and by about 0.1 mag in V, B,
and U compared with the 2005 October to 2006 January and 2006 September/October
observations followed by a rebrightening in the  2007 January observation.
This variability can be caused either by a change in the absorption column density
and therefore the reddening in the UV, or  by flux variations of the central
engine.
We also noticed that the flux in the UVOT filters agree with earlier measurements 
by the
International Ultraviolet Explorer taken between 1993-1995, but spectra taken by  
the Hubble Space Telescope Faint Object Spectrograph
show that WPVS 007 was fainter in the UV by a factor of at
least 2 in 1996. The flat optical/UV spectrum suggests that
some UV extinction is present
in the spectrum, but that alone cannot at all
account for the dramatic fading in the X-ray flux. 
 Most likely we see a
partial covering absorber in X-rays. Alternatively, the current X-ray emission
seen from WPVS 007 may also be the emission from the host galaxy.

\end{abstract}


\section{Introduction}

When observed at optical or lower energies, radio-quiet AGN appear to
be rather stable and not highly variable.  However, this picture
changes dramatically when AGN are observed in X-rays.  
Flux variability by factors 2-3 on timescales
of days to months is quite common among low- and
high-luminosity AGN. An increasing number that vary
by factors 10-30 has emerged
in the last decade, including in intermediate-type
Seyferts and Narrow-Line Seyfert 1 galaxies (NLS1s), and there is good
evidence that a substantial part of that variability
is caused by (cold) absorption, in terms of complete
or partial covering.
Interestingly, the highest amplitudes of variability
have been detected from the cores of non-active galaxies
in terms of transient flares, interpreted as tidal disruptions
of stars by the black holes at the centers of these galaxies
\citep[e.g.][ and references therein]{komossa04}.

NLS1s are objects with extreme properties: they show the steepest 
X-ray spectra,
strongest optical FeII emission and weakest [OIII] emission \citep[e.g.
][]{bol96, bor92, laor97, gru04a}.
The most
common explanation for their extreme properties is that they have
relatively small black hole mass and a high Eddington ratio $L/L_{\rm edd}$
\citep[e.g. ][]{bor02, gru04a, sul00}. 
NLS1s are also 
known to be objects with very strong X-ray variability \citep[e.g. ][]{lei99a}.

The NLS1 \wpvs\ 
\citep[1RXS J003916.6 $-511701$, RBS 0088;
$\alpha_{2000}$ = $00^{\rm h} 39^{\rm m} 15.^{\rm s}8$, 
$\delta_{2000}$ = $-51^{\circ} 17' 03 \farcs 0$, z=0.029; ][]{grupe95}
was discovered in a survey of faint southern galaxies with H$\alpha$ emission by
\citet{wamsteker85} and 
is a unique X-ray transient AGN. While X-ray transience is typically
associated with an X-ray outburst caused by a dramatic increase in the accretion
rate or the very onset of accretion,
the situation is very different in  WPVS 007.
During the ROSAT All-Sky Survey \citep[RASS, ][]{voges99},
   when the source was X-ray bright, 
   the optical-to-X-ray flux ratio was in a
   normal range for AGN \citep[e.g. ][]{beu99, maccacaro88}. 
   However, all follow-up X-ray observations between 1994 and 2002 
   using ROSAT and \chandra\  found it to have almost vanished 
   from the X-ray sky \citep{gru01, vau04}.

   Until recently the cause for the transient behavior of 
WPVS 007 had been a mystery. 
\citet{grupe95} suggested that this transience could be due to a
temperature change in the accretion disk that would shift the soft X-ray
spectrum out of the ROSAT PSPC energy observing window.
However, in recent years it became clear that the cause of the
transience is absorption. In 1996 July, WPVS 007 was observed by HST
\citep{goodrich00, constantin03}; a 2003 {\it FUSE} observation 
revealed the emergence of a  BAL flow \citep{leighly05, leighly06}. 
A discovery of a re-brightening and following rise in the X-ray
   luminosity, and in spectral changes, will set tight constraints on the
   movement and location of the absorber, and on the nature of the absorption.
   The high amplitude of the variability makes WPVS 007 exceptional among the
   known cases of absorption variability.
In order to detect a possible re-brightening and therefore the
disappearance of the absorber, we began a monitoring campaign using \swift\ in
2005 October.

The \swift\ mission \citep{gehrels04}
was launched on 2004 November 20th. The main purpose is to
hunt and observe Gamma-Ray Bursts (GRBs). 
However, part of the observing time is used for fill-in
targets and targets-of-opportunity
when no GRB can be observed. Due to its multi-wavelength capacities and
its flexible scheduling, \swift\ is an ideal observatory of all types of 
AGN, as
 demonstrated by e.g \citet{grupe06} on the NLS1 RX J0148.3--2758 and 
\citet{markwardt05} on the search for obscured AGN in the BAT survey.
\swift\ is equipped with three telescopes: 
at the high energy end the Burst Alert
Telescope \citep[BAT, ][]{barthelmy04} operating in the 15-150 keV 
energy range,  
the X-Ray Telescope 
\citep[XRT, ][]{burrows04}, which covers the soft X-ray range between 0.3-10.0
keV, and at the long wavelength end, the UV-Optical
Telescope \citep[UVOT, ][]{roming04}.
The XRT uses a CCD detector identical to the EPIC MOS on-board XMM \citep{tur01}. 
The UVOT covers the range between 1700-6500\AA and
 is a sister instrument
 of XMM's Optical Monitor
\citep[OM, ][]{mason01}. The UVOT has a similar set of filters as the OM
\citep{mason01, roming04}. However, the UVOT UV throughput is a factor of about
10 higher than in the OM.

The outline of this paper is as follows: in \S\,\ref{observe} we describe the
\swift\ 
observations and the data reduction, in \S\,\ref{results} we
present the results of the \swift\ XRT and UVOT data analysis and compare
the UVOT
data with earlier IUE and HST spectra, 
and in \S\,\ref{discuss} we discuss the results. 
Throughout the paper spectral indexes are denoted as energy spectral indexes
with
$F_{\nu} \propto \nu^{-\alpha}$. Luminosities are calculated assuming a $\Lambda$CDM
cosmology with $\Omega_{\rm M}$=0.27, $\Omega_{\Lambda}$=0.73 and a Hubble
constant of $H_0$=75 km s$^{-1}$ Mpc$^{-1}$ using a luminosity distances D=118 Mpc
given by \citet{hogg99}. All errors are 1$\sigma$ unless stated otherwise.

\section{\label{observe} Observations and data reduction}

\wpvs\ has been monitored
 by \swift\ between 2005 October and 2007 January. 
Table\,\ref{xrt_log} lists the \swift~XRT observations, including the start and
end times, the total exposure times, and the 3$\sigma$ upper limits. Note that we do not include
the segment 005 data (2006 September 06) in the XRT analysis because during the 
time of that
observation the XRT detector was rather warm resulting in an enhanced detector background. However,
the UVOT data from that time period were not affected.
The \swift~UVOT observations are summarized in Table\,\ref{uvot_log}. 
Also note that segment 008 does not exist. WPVS 007 was originally scheduled
 for 2006 December 03, but
the observations were superseded by the detections of GRBs
 061201 and 061202 \citep[][respectively]{marshall06, sakamoto06}
 before the start of the WPVS 007
observations. Segment numbers, however, can only be used once by the \swift\ scheduling tool.
The
segment numbers in both table refer to the days \swift\ observed WPVS 007
 \citep[See the description in ][]{grupe06}.
In the first observation of 2006 December (segment 009) we noticed that the AGN
became significantly fainter in the UV by 0.2 mag. In order to investigate this
behavior and to get a better estimate of the time scale we initiated an
additional ToO for two pointings of 2 ks each which were executed on 2006
December 12 and 21 (segments 010 and 011). Also note that the 2007 January
observation was split into two segments due to scheduling constrains, even
though the observations were performed in consecutive orbits. 

The XRT was operating in photon counting mode \citep{hill04} and the
data were reduced by the task {\it xrtpipeline} version 0.10.4., which is
included in the HEASOFT package 6.1. Photons were collected in the 0.3-10.0
keV energy range.
The upper
limits were determined from the background in the XRT. The photons were
extracted with {\it XSELECT} version 2.4. In order to compare the observations
from different missions we use the HEASARC tool {\it PIMMS} version 3.8. For the
conversion we assumed an absorbed power law model with the absorption column
density at the Galactic value \citep[2.84 $\times$ 10$^{20}$ cm$^{-2}$,
][]{dic90} and an energy spectra slope \ax=3.0. Note, however, that this is just
an estimate, since we do not know what the low-state X-ray spectrum of \wpvs\ really
looks like. For all observations the counts were corrected due to the exposure maps.

As listed in Table\,\ref{uvot_log} UVOT observations were performed in all filters
except during the 2005 October and December, and 2006 January observations only the UV
filters were used. 
Photometry on all UVOT individual and coadded exposures was performed with 
the tool {\it uvotmaghist} version 0.1. A $6\arcsec$ and 
$12\arcsec$ radius extraction regions  
were used centered on WPVS 007 for the optical (V, B, and U) and UV filters (UV W1, UV
M2, and UV W2), respectively
and the background count rate was measured
 with a $20\arcsec$ radius aperture in a nearby source free region. 
 All reported magnitudes have been corrected for Galactic extinction, 
 where the reddening in the line-of-sight to the object is 
 $E(\bv) = 0.012$~mag~\citep{sfd98}.
All data were aspect corrected
and coadded before measuring the magnitudes and fluxes.
The UVOT uses Vega-based magnitudes with the following zeropoints:
V=17.88\plm0.09, B=19.16\plm0.12, U=18.38\plm0.23, UVW1=17.69\plm0.20,
UVM2=17.29\plm0.23, and UVW2=17.77\plm0.20 \citep{brown07}.

Prior the \swift\ observations WPVS 007 was observed in the UV by IUE four times
between 1993 and 1995, HST in 1996 July, and FUSE in 2003 November. 
In this paper we make
use of the IUE and HST data. Table\,\ref{iue_hst_log} lists these observations.
The FUSE data will be presented by
\citet{leighly06} which will also contain a spectral
  analysis of the miniBALs present in the HST data.

\section{\label{results} Results}

\subsection{X-rays}

We do not detect WPVS 007 in X-rays in any of the 
monitoring observations performed by
\swift\ so far as listed in Table\,\ref{xrt_log}.
To determine 3$\sigma$ upper limits we applied the method by \citet{kraft91}. This
method determines the confidence levels for low numbers of counts using the
Bayesian method for Poisson-distributed data. The background in all three
observations was measured in a circular region with r=235$^{''}$ around the
position of \wpvs. For the source itself we assumed an extraction radius of
23.5$^{''}$. 
The 3$\sigma$
upper limits are listed in Table\,\ref{xrt_log}. 
We coadded all XRT observations together, except for the 2006 September 06 observation
(segment 005) when the background was too high. 
From these coadded data with a total exposure time of 23.2 ks we measured an
upper limit of 1.04$\times 10^{-3}$ counts s$^{-1}$ in the \swift-XRT. Assuming a power
law model spectrum with \ax=3.0 and \nh\ at the Galactic value this upper limit count
rate converts to an upper limit in unabsorbed flux in the 0.3-10 keV band of
$2.6\times 10^{-17}$ W m$^{-2}$. 

\begin{figure*}
\epsscale{1.2}
\plotone{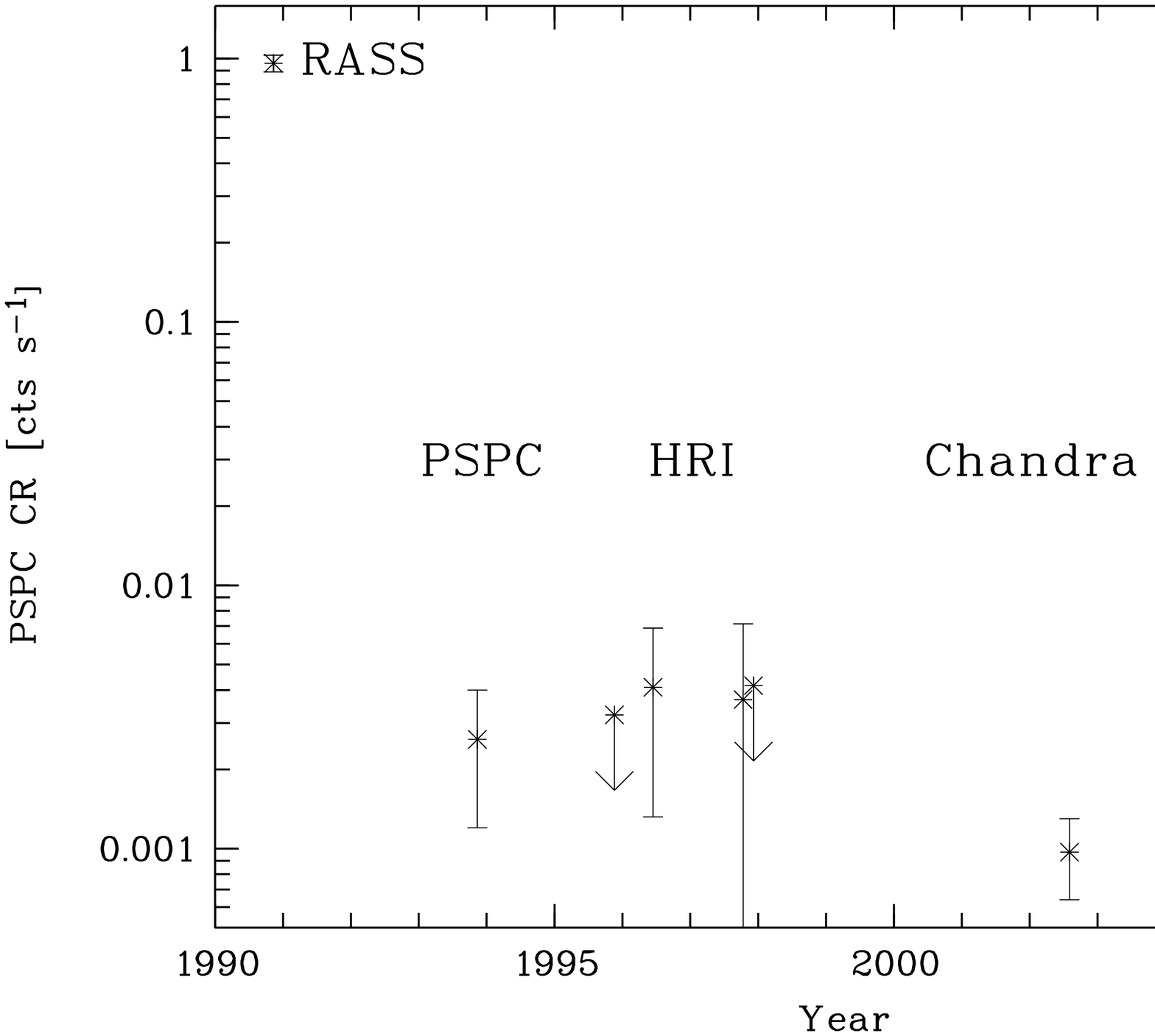}
\caption{\label{wpvs007_lc_all} Long-term light curve of \wpvs\ containing the
ROSAT All-Sky Survey and pointed PSPC and HRI observations, the \chandra, and
the upper limits derived from the \swift~XRT observations. The count rates were
converted by assuming an absorbed power law model with \nh=2.84$\times 10^{20}$
cm$^{-2}$ and \ax=3.0. The large red downward arrow displays the upper limit of the
coadded exposures of the \swift-XRT observations (23.2 ks).
}
\end{figure*}

Figure\,\ref{wpvs007_lc_all} displays the long term light curve of \wpvs. The
ROSAT values were taken from \citet{gru01} and the \chandra\ data point from
\citet{vau04}. The light curve shows that \wpvs\ is still in a low state and
 the upper limit of the coadded \swift\ observations is consistent with the 
ROSAT PSPC and HRI detections. Note that Figure\,\ref{wpvs007_lc_all} uses PSPC
counts which were converted by PIMMS assuming a power law spectrum with \ax=3.0
and \nh\ at the Galactic value.

\subsection{UVOT Photometry}

\begin{figure*}
\epsscale{1.0}
\plotone{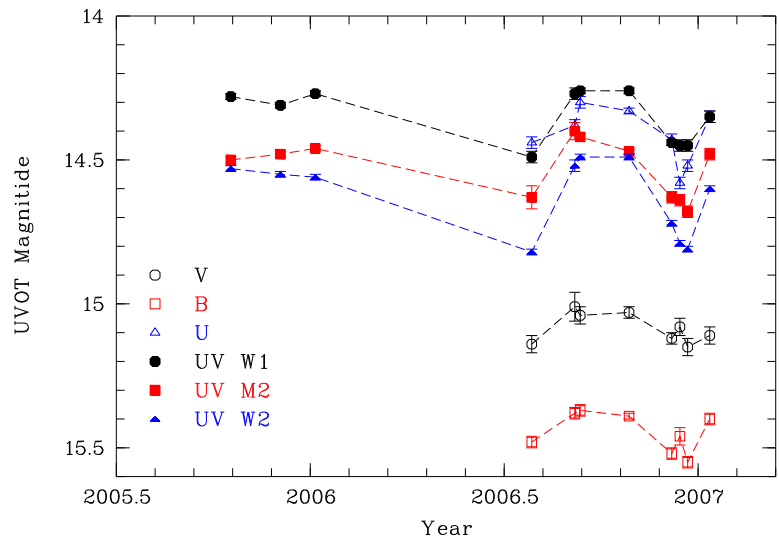}
\caption{\label{wpvs007_uvot_lc} \swift~UVOT light curves of WPVS 007. The
values are given in Table\,\ref{uvot_log}.
}
\end{figure*}

The magnitudes in the UVOT filters are given in Table\,\ref{uvot_log} and
previous UV observations by IUE and HST are listed in Table\,\ref{iue_hst_log}.
The errors quoted in Table\,\ref{uvot_log} are statistical errors. 
As listed in
Table\,\ref{uvot_log} and shown in Figure\,\ref{wpvs007_uvot_lc}
WPVS 007 became fainter in the optical filters by about 0.1 mag
and  by about 0.2 mag in the UV filters in the 2006 July and December observations
compared with the observations obtained in 2005 and 2006 January and 2006 September
and October. The AGN became brighter again by about 0.2 mag in the 2007
January observation which was about three weeks after the last observation in
2006 December.
From the current light curve as shown in Figure\,\ref{wpvs007_uvot_lc}
the AGN seems to be variable in the optical/UV on timescales 
of a few weeks to several months.   

\begin{figure*}
\epsscale{1.0}
\plotone{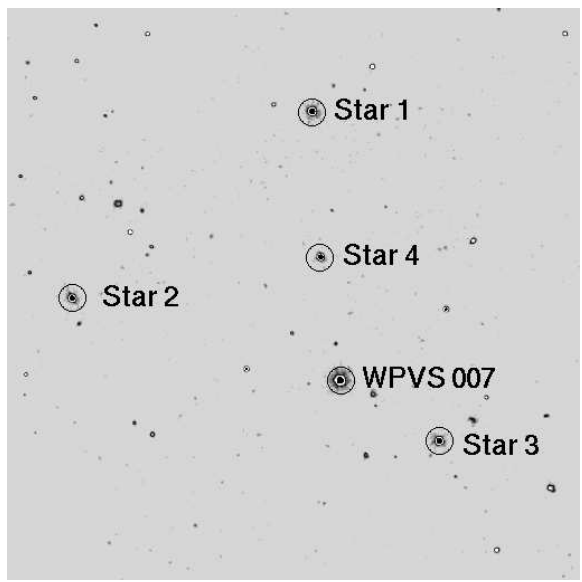}
\caption{\label{wpvs007_uvw2_image} \swift~UVOT W2 image of the 2006 January
observation with the reference stars as listed in Table\,\ref{ref_stars} and
WPVS 007.
}
\end{figure*}

In order to determine whether this
variability is real or just within the uncertainties we picked 4 field
stars of similar magnitude as WPVS 007 as reference stars and compared their
magnitudes segment by segment.
Table\,\ref{ref_stars} lists these 4 stars with the coordinates and their
magnitudes in U, UV W1, UV M2, and UV W2 with statistical errors.
Figure\,\ref{wpvs007_uvw2_image} displays
where these stars are located relative to WPVS 007. Note that during some of the
observations not all the 4 stars were in the field of view of the UVOT.
U magnitudes could only be given for Stars \#3 and 4, because \#1 and 2 are too bright in
the U filter that they suffer significantly from coincident losses. Also note
that star \#1 and \#2 were not in the field of view during some of the
observations.
As shown in
Table\,\ref{ref_stars}, the variance in the field stars is small compared with the
variability observed in WPVS 007.
Therefore, we consider the UV variability found in
WPVS 007 to be real. The change by 0.2 mag in the UV filters observed during the 2006
July and December observations is larger than the uncertainties between the measurements
in the field stars.

\begin{figure*}
\epsscale{1.0}
\plotone{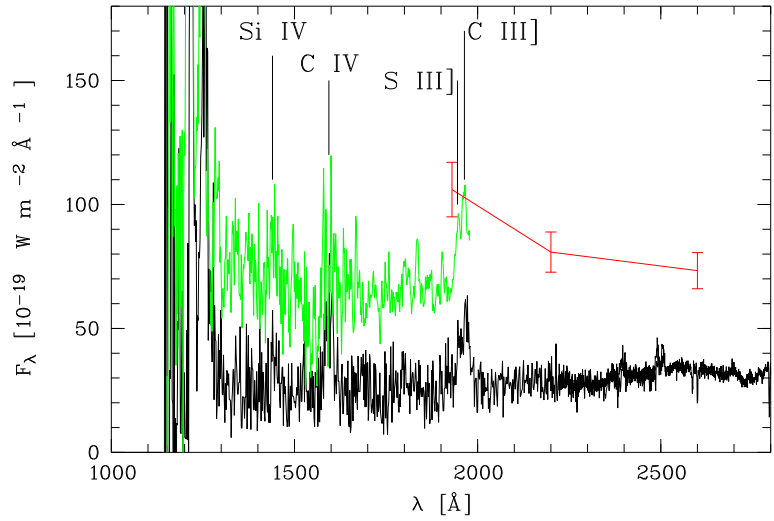}
\caption{\label{wpvs_uv_spec} UV observations of \wpvs. The red points are the
\swift\ UVOT observations from 2006 January,
the black spectrum is
the HST observation from 1996, and the green spectrum the average of the IUE observation from
1993, 1994, and 1995 December.
}
\end{figure*}

WPVS 007 has shown variability on timescales of months before between the
IUE and HST observations between 1993 and 1996.
Figure\,\ref{wpvs_uv_spec} displays the \swift\ UVOT measurements
from 2006 January, the HST spectrum from 1996 \citep{goodrich00, constantin03}, 
and the IUE spectrum averaging three spectra taken between 1993 and 1995. We
excluded the 1995 November observation due to strong cosmic ray events during
that observation. The data shown in Figure\,\ref{wpvs_uv_spec} are the
observed data, uncorrected for reddening.
The IUE spectra and the \swift-UVOT data seem to agree. However, the HST spectra
show a significantly lower flux than the IUE and \swift\ data. \citet{dunn06}
presented an Internet database\footnote{http://www.chara.gsu.edu/PEGA/IUE}
of UV continuum light curves of Seyfert galaxies
which also showed that WPVS 007 displays 
significant variability in the UV between the IUE and HST observations.
One possibility of a lower flux in the HST data is a mis-alignment of the source in the
$1\arcsec$ aperture in the HST FOS. However, as listed in Table\,\ref{iue_hst_log}, the
AGN was observed during several orbits and the fluxes in all these spectra agree with
each other. A $1\arcsec$ aperture is also rather large compared to the $0\farcs1$
resolution. Therefore we exclude a mis-allignment as the cause of the lower flux in WPVS
007 during the HST observation in 1996 July. 
Note that \citet{winkler92} reports
V=15.28\plm0.03, B=15.77\plm0.03, and U=15.15\plm0.03. While the small
differences in V and B can be explained by the different central wavelength
between the filters used by \citet{winkler92} and the UVOT, the difference in U
suggests that WPVS 007 is variable in U.

\subsection{Spectral Energy Distribution}

\begin{figure*}
\epsscale{1.0}
\plotone{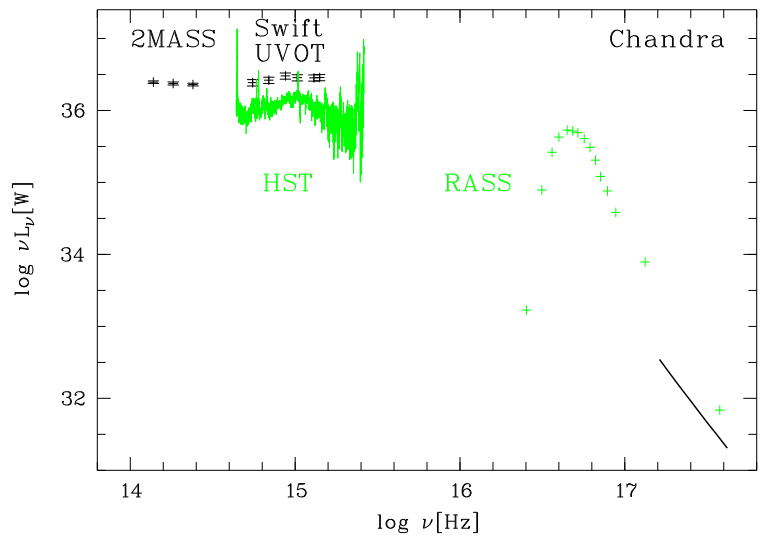}
\caption{\label{wpvs007_sed} Spectral energy distribution of WPVS 007 using the NIR
data from 2MASS, optical/UV from the \swift~UVOT observation from January 2006
corrected for Galactic reddening, 
and the X-ray data from the
\chandra\ observation from 2002. 
The Chandra data are displayed as a power law model
with \ax=3.0.
Note that all these observations have not been
performed simultaneously. 
The HST spectra and the 
observed RASS data are
displayed in light green/grey for comparison purposes.
}
\end{figure*}

Figure\,\ref{wpvs007_sed} displays the spectral energy distribution of WPVS 007
using 2MASS NIR data derived from the Nasa Extragalactic database, optical/UV
data from \swift's UVOT from 2006 July
and the X-ray data from the \chandra\ observation from
2002 \citep{vau04}. The Chandra data are displayed as an un-absorbed power
law model with \ax=3.0.
The optical-to-Xray slope \aox\footnote{The
X-ray loudness is defined by \citet{tananbaum79} as \aox=--0.384
log($f_{\rm 2keV}/f_{2500\AA}$).} measured from this plot is \aox=5.4. This is
an extreme value for an AGN which typically have values around \aox=1.5
\citep[e.g. ][]{yuan98a, strateva05}. Assuming an \aox=1.5 we would expect a
flux at 2 keV $F_{\rm 2keV}=6\times10^{-16}$ W m$^{-2}$, or a luminosity at 2
keV $L_{\rm 2 keV}=1\times10^{35}$ W. However, as shown in
Figure\,\ref{wpvs007_sed}, this is not the case. During the RASS observation,
assuming an UV spectrum like during the \swift\ observations, the \aox was in
the order of \aox=5.0. 
The  flattening of the UV 
spectrum, as
shown by the UVOT photometry data, suggests some intrinsic
reddening of the AGN.
A NLS1 typically has a very blue optical/UV spectrum, as shown e.g in
\citet{grupe06} for the NLS1 RX J0148.3-2758. Assuming that the intrinsic optical
UV spectrum of \wpvs\ is similar to that of RX J0148.3-2758, we can estimate a
reddening by 0.6 mag in the UV W2 filter. This results in a lower limit of the 
intrinsic reddening $\rm E_{B-V}$=0.073. Note that this is a rough estimate.
Based on the H$\alpha$/H$\beta$ flux ratio,
\citet{winkler92} estimated the extinction to $A_{\rm V}\approx$1.0 which is
significantly higher than our estimate.

\section{\label{discuss} Discussion}

Our main results are that \wpvs\ is still in a low state in X-rays and that it
shows significant variability in the UV on timescales of months.
Adding all
 \swift\ observations together (except for the segment 005 data of 2006 September 06)
 we determine a 3$\sigma$ upper limit =
1.04$\times10^{-3}$ counts s$^{-1}$, which converts to an unabsorbed flux in the 0.3-10
keV band of $2.6\times 10^{-17}$ W m$^{-2}$.
This upper limit is at a similar level to the
ROSAT pointed PSPC and HRI observations. Note that Chandra, with its superb point
spread function, was able to detect WPVS 007 at an even lower level
\citep{vau04} within an exposure time of 10 ks.
Even though we could not detect a re-brightening of WPVS 007
in X-rays, it is still an exciting AGN. Note that the purpose of the \swift\
observations is not to obtain a deep detection, but to monitor the AGN in order to
detect when it re-brightens again. 

Our UVOT observations suggest that the AGN is variable 
in the UV bands by about 0.2 mag
within timescales of a few months.
 UV variability,
however, is not uncommon in AGN and has been reported for various AGN such as
NGC 4151 \citep{edelson96, crenshaw96}, NGC 5548 \citep{clavel91, korista95},
Akn 564 \citep{collier01}, and 3C 390.3 \citep{obrien98}, but only a few NLS1s
have repeated UV coverage as good as WPVS 007.
In the case of WPVS
007 we can speculate that the UV variability is likely to be caused by a change
in the absorption column density and therefore the reddening in the UV. 
Using the change in
the UV W1 magnitude by 0.2 mag we can derive an additional $E_{\rm B-V}$=0.032. This
would cause an additional reddening by 0.10 mag in V, 0.17 in U, and 0.28 in UV W2 which
is consistent with the changes we observe during the 2006 July and December observations.
However, a change in luminosity of the central engine cannot completely been excluded
based on the current data set.

The variability we observed in the UV between the IUE,  HST and \swift-UVOT
observations was previously also noticed by \citet{dunn06} between the IUE and
HST observations. This is not a calibration issue. 
There are no problems reported on the HST FOS and IUE
observations either. Also, looking at the light curves presented in the UV
continuum light curve database by \citet{dunn06} suggest that the UV flux is
already decaying between the IUE observation of 1994 to 1995 and that the 1996
HST observation is consistent with this decay.  
 
WPVS 007 may be one of the most extreme cases of X-ray weak NLS1s such as those found by
\citet{williams02, williams04}. The reddened optical/UV spectrum also suggests that
the X-ray weakness of WPVS 007 is caused by absorption. However, of the 10
photons detected at the position of WPVS 007 in the \chandra~ACIS-S data
\citep{vau04}, 8 have energies below 1 keV. A simple cold absorber would have
absorbed all these photons. The solution could be a partial covering absorber that
would allow some of the soft X-ray photons to escape. Such partial covering 
absorbers
have been found in several NLS1s, such as Mkn 1239 \citet{grupe04c} or 1H
0707-495 \citet{gallo04, tanaka04, boller02}.

Alternatively, the low-state X-ray emission detected with
  Chandra may actually represent the X-ray emission from the host
  galaxy; the AGN being completely absorbed.
  In order to check this, we used the blue magnitude of the
  galaxy to predict the expected X-ray flux from the host galaxy,
  using the correlation between $L_{\rm B}$ and $L_{\rm X}$ for early-type  
  galaxies of \citet{osullivan01}. We use the extinction-
  corrected blue magnitudes measured with \swift, $m_{\rm B,corr}$=15.48, 
  and a previous USNO measurement, $m_{\rm Bcorr}$=14.24.{\footnote{
  Both measurements have to be taken with some caution, since 
  we did not correct for any possible AGN contribution to $m_{\rm B}$
  on the one hand, and since the short \swift\ observation may
  have missed part of the host galaxy contribution on the other
  hand. Therefore, the estimate should only be regarded as 
  order of magnitude}}. We then predict a host galaxy
  contribution to the X-ray luminosity of $L_{\rm X} \approx 10^{33} - 
  7 \times 10^{33}$ W,
  which may well account for all of the observed low-state
  emission. Indeed, host galaxies typically show such
               soft, thermal spectra. 
  However, as shown in Figure 1 in \citet{vau04} the
  10 photons found at the position of WPVS 007 seem to be consistent with a
  point source. In order to verify this statement, a much longer observation by
  \chandra\ is needed.
We will continue monitoring WPVS 007 every 4-6 weeks
with \swift\ as long as the AGN is not sun-constrained. Especially more observations in
the UV will give us a better handle on the timescales of the variability in the UV.

\acknowledgments

First we want to thank Neil Gehrels for approving our ToO requests and
the \swift\ team for performing the ToO observations of
WPVS 007 and scheduling the AGN on a regular basis.
We would also like to thank Jay Dunn for quickly checking the IUE and HST data for
any problems,  Matthias Dietrich for various discussions about UV variability
in AGN, and the anonymous referee for useful comments and suggestions to
improve the paper.
This research has made use of the NASA/IPAC Extragalactic
Database (NED) which is operated by the Jet Propulsion Laboratory,
Caltech, under contract with the National Aeronautics and Space
Administration.
 This research was supported by NASA contract NAS5-00136 (D.G.,  \& J.N.).

\clearpage



\clearpage

\begin{deluxetable}{lccrc}
\tablecaption{\swift~XRT Observation log of WPVS 007
\label{xrt_log}}
\tablewidth{0pt}
\tablehead{
\colhead{Segment} & \colhead{T-start\tablenotemark{1}} & 
\colhead{T-stop\tablenotemark{1}} &
\colhead{$\rm T_{exp}$\tablenotemark{2}} & \colhead{3$\sigma$
ul\tablenotemark{3}}
} 
\startdata
001 & 2005-10-20 03:21 & 2005-10-20 03:54 & 2025 & 5.78 \\
002 & 2005-12-07 13:29 & 2005-12-07 13:58 & 1696 & 3.42 \\
003 & 2006-01-05 00:48 & 2006-01-05 05:41 & 3466 & 1.67 \\
004 & 2006-07-31 04:39 & 2006-07-31 06:30 & 1610 & 6.16 \\
005 & 2006-09-06 08:19 & 2006-09-06 11:38 &  917 & ---\tablenotemark{4} \\
006 & 2006-09-11 05:40 & 2006-09-11 07:37 & 2335 & 5.02 \\
007 & 2006-10-27 02:34 & 2006-10-27 23:33 & 3228 & 1.80 \\
009 & 2006-12-06 00:02 & 2006-12-06 06:36 & 2964 & 2.70 \\
010 & 2006-12-12 00:33 & 2006-12-12 01:01 & 1611 & 3.60 \\
011 & 2006-12-21 06:19 & 2006-12-21 08:15 & 2273 & 4.36 \\
012+013 & 2007-01-10 21:21 & 2007-01-11 00:53 & 1905 & 3.05 \\
001-013\tablenotemark{4} & 2005-10-20 03:21 & 2007-01-11 05:53 
 & 23198\tablenotemark{4} & 1.04\tablenotemark{4} \\
\enddata

\tablenotetext{1}{Start and End times are given in UT}
\tablenotetext{2}{Observing time given in s}
\tablenotetext{3}{3$\sigma$ upper limits in units of $10^{-3}$ XRT counts s$^{-1}$}
\tablenotetext{4}{Due high background during the observation the segment 005 XRT data were not
included in the analysis. Segment 008 does not exist (see text)}
\end{deluxetable}

\begin{deluxetable}{lrcrcrcrcrcrc}
\tabletypesize{\tiny}
\tablecaption{\swift~UVOT Observation of WPVS 007
\label{uvot_log}}
\tablewidth{0pt}
\tablehead{
& \multicolumn{2}{c}{V} 
& \multicolumn{2}{c}{B} 
& \multicolumn{2}{c}{U}  
& \multicolumn{2}{c}{UV W1} 
& \multicolumn{2}{c}{UV M2} 
& \multicolumn{2}{c}{UV W2}  \\
\colhead{\rb{Segment}} & 
\colhead{$\rm T_{exp}$\tablenotemark{1}} & \colhead{$\rm Mag_{corr}$\tablenotemark{2}} &
\colhead{$\rm T_{exp}$\tablenotemark{1}} & \colhead{$\rm Mag_{corr}$\tablenotemark{2}} &
\colhead{$\rm T_{exp}$\tablenotemark{1}} & \colhead{$\rm Mag_{corr}$\tablenotemark{2}} &
\colhead{$\rm T_{exp}$\tablenotemark{1}} & \colhead{$\rm Mag_{corr}$\tablenotemark{2}} &
\colhead{$\rm T_{exp}$\tablenotemark{1}} & \colhead{$\rm Mag_{corr}$\tablenotemark{2}} &
\colhead{$\rm T_{exp}$\tablenotemark{1}} & \colhead{$\rm Mag_{corr}$\tablenotemark{2}} 
} 
\startdata
001 & \nodata & \nodata & \nodata & \nodata & \nodata & \nodata & 656    & 14.28\plm0.01 & 686  & 14.50\plm0.01 & 686  & 14.53\plm0.01 \\
002 & \nodata & \nodata & \nodata & \nodata & \nodata & \nodata & 550    & 14.31\plm0.01 & 588  & 14.48\plm0.01 & 588  & 14.55\plm0.01 \\
003 & \nodata & \nodata & \nodata & \nodata & \nodata & \nodata & 1056   & 14.27\plm0.01 & 1171 & 14.46\plm0.01 & 1171 &  14.56\plm0.01 \\
004 & 155 & 15.14\plm0.03 & 159 & 15.48\plm0.02 & 159 & 14.44\plm0.02 & 319  & 14.49\plm0.02 &  118 & 14.63\plm0.04 & 615  & 14.82\plm0.01 \\
005 &  55 & 15.01\plm0.05 & 170 & 15.38\plm0.02 & 170 & 14.38\plm0.02 & 340  & 14.27\plm0.02 &  144 & 14.40\plm0.03 & 392  & 14.52\plm0.02 \\
006 & 194 & 15.04\plm0.03 & 194 & 15.37\plm0.02 & 194 & 14.30\plm0.02 & 387  & 14.26\plm0.01 &  536 & 14.42\plm0.01 & 777  & 14.49\plm0.01 \\
007 & 344 & 15.03\plm0.02 & 336 & 15.39\plm0.01 & 335 & 14.33\plm0.01 & 686  & 14.26\plm0.01 &  767 & 14.47\plm0.01 & 767  & 14.49\plm0.01 \\
009 & 245 & 15.12\plm0.02 & 245 & 15.52\plm0.02 & 245 & 14.43\plm0.02 & 486  & 14.44\plm0.01 &  621 & 14.63\plm0.02 & 978  & 14.72\plm0.01 \\
010 & 134 & 15.08\plm0.03 & 134 & 15.46\plm0.03 & 134 & 14.58\plm0.02 & 267  & 14.45\plm0.01 &  376 & 14.64\plm0.02 & 534  & 14.79\plm0.01 \\
011 & 187 & 15.15\plm0.02 & 187 & 15.55\plm0.02 & 187 & 14.52\plm0.02 & 374  & 14.45\plm0.02 &  513 & 14.68\plm0.02 & 750  & 14.81\plm0.01 \\
012+013 & 155 & 15.11\plm0.03 & 155 & 15.40\plm0.02 & 155 & 14.35\plm0.02 & 312 & 14.35\plm002 & 373 & 14.48\plm0.02 & 625 & 14.60\plm0.01\\
\enddata
\tablenotetext{1}{Observing time given in s}
\tablenotetext{2}{Magnitude corrected for reddening with $E_{\rm B-V}$=0.012 given by
\citet{sfd98}. The errors given in this table are statistical errors}
\end{deluxetable}

\begin{deluxetable}{ccccr}
\tablecaption{Previous UV observations of WPVS 007
\label{iue_hst_log}}
\tablewidth{0pt}
\tablehead{
\colhead{Mission} & \colhead{ObsID} & 
\colhead{grating} &
\colhead{T-start\tablenotemark{1}} & 
\colhead{$\rm T_{exp}$\tablenotemark{2}} 
} 
\startdata
IUE & SWP 48542 & --- & 1993-09-05 15:58 & 16800 \\
    & SWP 52369 & --- & 1994-10-10 15:30 & 18900 \\
    & SWP 56215 & --- & 1995-11-19 22:00 & 20000 \\
    & SWP 56318 & --- & 1995-12-20 12:47 & 18000 \\
HST\tablenotemark{3} & FOS Y3790102T & G130H & 1996-07-30 13:30 & 1730 \\
                     & FOS Y3790103T & G130H & 1996-07-30 14:46 & 2110 \\
		     & FOS Y3790104T & G160L & 1996-07-30 15:30 &  240 \\
		     & FOS Y3790105T & G190H & 1996-07-30 16:29 & 1500 \\
		     & FOS Y3790107T & G270H & 1996-07-30 18:02 & 1280  \\ 
\enddata

\tablenotetext{1}{Start and End times are given in UT}
\tablenotetext{2}{Observing time given in s}
\tablenotetext{3}{A complete listing of all HST FOS observations is given in
\citet{constantin03}.}
\end{deluxetable}

\begin{deluxetable}{cccccccccccc}
\tabletypesize{\tiny}
\rotate
\tablecaption{List of reference stars used to determine the error in the
photometry. The magnitudes are the uncorrected, directly measured values.
The positions relative to WPVS 007 are shown in
Figure\,\ref{wpvs007_uvw2_image}.
\label{ref_stars}}
\tablewidth{0pt}
\tablehead{
\colhead{Object} & \colhead{$\alpha_{2000}$} & 
\colhead{$\delta_{2000}$} & 
\colhead{Filter} & \colhead{Segment 003} & \colhead{Segment 004}
& \colhead{Segment 006} & \colhead{Segment 007}
& \colhead{Segment 009} & \colhead{Segment 010} & \colhead{Segment 011} 
& \colhead{Seg. 012+013}
} 
\startdata
Star 1 & 00 39 20.0 & -51 10 53.2 & UV W1 & 13.50\plm0.01 & \nodata       & \nodata       & 13.50\plm0.01 & 13.51\plm0.01 & 13.53\plm0.01 & 13.52\plm0.01 & 13.50\plm0.01 \\
       &            &             & UV M2 & 16.27\plm0.02 & \nodata       & \nodata       & 16.26\plm0.03 & 16.24\plm0.03 & 16.30\plm0.04 & 16.20\plm0.03 & 16.22\plm0.04 \\
       &            &             & UV W2 & 15.32\plm0.02 & \nodata       & \nodata       & 15.26\plm0.01 & 15.02\plm0.01 & 15.10\plm0.02 & 15.29\plm0.01 & 15.32\plm0.02 \\
Star 2 & 00 39 54.2 & -51 15 04.7 & UV W1 & 14.38\plm0.01 & 14.40\plm0.02 & 14.40\plm0.01 & \nodata       & \nodata       & 14.39\plm0.02 & 14.44\plm0.02 & \nodata \\
       &            &             & UV M2 & 15.77\plm0.02 & 15.78\plm0.06 & 15.80\plm0.03 & \nodata       & \nodata       & 15.79\plm0.03 & 15.77\plm0.03 & \nodata \\
       &            &             & UV W2 & 15.93\plm0.02 & 15.98\plm0.02 & 15.95\plm0.02 & \nodata       & \nodata       & 15.98\plm0.02 & 15.99\plm0.02 & \nodata \\         
Star 3 & 00 39 01.8 & -51 18 17.6 & U     & \nodata       & 13.21\plm0.02 & 13.27\plm0.02 & 13.26\plm0.01 & 13.21\plm0.02 & 13.29\plm0.02 & 13.25\plm0.02 & 13.21\plm0.02 \\
       &            &             & UV W1 & 14.30\plm0.01 & 14.32\plm0.02 & 14.37\plm0.01 & 14.33\plm0.01 & 14.30\plm0.01 & 14.35\plm0.02 & 14.33\plm0.02 & 14.32\plm0.02 \\
       &            &             & UV M2 & 15.25\plm0.01 & 15.24\plm0.04 & 15.35\plm0.02 & 15.36\plm0.02 & 15.29\plm0.01 & 15.38\plm0.03 & 15.31\plm0.02 & 15.27\plm0.02 \\
       &            &             & UV W2 & 15.50\plm0.01 & 15.55\plm0.02 & 15.65\plm0.02 & 15.54\plm0.01 & 15.55\plm0.01 & 15.64\plm0.02 & 15.53\plm0.02 & 15.54\plm0.02 \\
Star 4 & 00 39 18.7 & -51 14 14.0 & U     & \nodata       & 13.89\plm0.02 & 13.87\plm0.02 & 13.88\plm0.02 & 13.93\plm0.02 & 13.92\plm0.02 & 13.96\plm0.02 & 13.91\plm0.02 \\
       &            &             & UV W1 & 15.06\plm0.01 & 15.04\plm0.02 & 15.04\plm0.02 & 15.04\plm0.01 & 15.08\plm0.02 & 15.05\plm0.02 & 15.04\plm0.02 & 15.09\plm0.03 \\
       &            &             & UV M2 & 16.17\plm0.02 & 16.07\plm0.06 & 16.08\plm0.03 & 16.12\plm0.02 & 16.07\plm0.03 & 16.14\plm0.04 & 16.14\plm0.03 & 16.07\plm0.03 \\
       &            &             & UV W2 & 16.50\plm0.02 & 16.41\plm0.03 & 16.43\plm0.02 & 16.46\plm0.02 & 16.48\plm0.02 & 16.43\plm0.03 & 16.47\plm0.03 & 16.55\plm0.03 \\
                             
\enddata

\end{deluxetable}

\end{document}